\documentclass[prd,preprint,tightenlines,floatfix,showpacs,preprintnumbers,nofootinbib,eqsecnum]{revtex4}

 \usepackage[dvips,final]{graphicx}
  \usepackage{amssymb}
   \usepackage{amsmath}
    \usepackage{amsfonts}
     \usepackage{epsfig}
      \usepackage{bm}

\newcommand{\ga}{\gamma}

\newcommand{\si}{\sigma}

\renewcommand{\epsilon}{\varepsilon}

\begin{document}

\thispagestyle{empty} \preprint{\hbox{}} \vspace*{-10mm}

\title{Renormgroup origin and analysis of Split Higgsino scenario}

\author{V.~A.~Beylin}

\email{vbey@rambler.ru}

\author{V.~I.~Kuksa}

\email{kuksa@list.ru}

\author{G.~M.~Vereshkov}

\email{gveresh@gmail.com}

\affiliation{Institute of Physics,
Southern Federal University (former Rostov State University),
Rostov-on-Don 344090, Russia}

\author{R.~S.~Pasechnik}

\email{rpasech@theor.jinr.ru}

\affiliation{Bogoliubov Laboratory of Theoretical Physics, JINR,
Dubna 141980, Russia}

\date{\today}

\begin{abstract}
We present a renormalization group motivation of scale hierarchies in
SUSY $SU(5)$ model. The Split Higgsino scenario with a high scale of the SUSY
breaking is considered in detail. Its manifestations in experiments
are discussed.
\end{abstract}

\pacs{12.60.Jv, 95.35.+d, 95.30.Cq}


\maketitle

\section{Introduction}

Supersymmetric generalization of the SM permits one to solve some
inner problems of the theory. However, the MSSM is an ambiguous
model: despite the defined set of physical fields, only one
characteristic MSSM parameter -- electroweak scale $M_{EW}$ -- has
been fixed in experiment. The scales of scalar quarks and leptons
$M_0$, gaugino $M_{1/2}$, and Higgsino $\mu$, can be picked out as
arbitrary ones. Expectations of new dynamical effects of the
supersymmetry at the LHC are induced in the MSSM by some specific
choice of the SUSY breaking scale that is not very far from the EW
scale, $M_{SUSY}\sim M_{1/2}\sim M_0\sim O\,(1\;\mathrm{TeV})$. This
scale hierarchy seems ``natural'', providing regularization of
quadratic divergencies long before $M_{GUT}$. Thus, the MSSM as it
is motivated by some theoretical and phenomenological arguments --
successful RG evolution, ``natural''\ choice of the renormalization
scale and the reasonable dark matter (DM) description -- works well
for the specific choice of the SUSY breaking scale only. However,
the ``naturalness''\ as one of the MSSM principles does not seem to
be the obligatory requirement  from the QFT point of view. So
establishing of the genuine SUSY scales hierarchy is a problem which
can be unambiguously solved beyond the MSSM.

The arbitrariness in the choice of the scale hierarchy can be
sufficiently diminished when the gauge coupling unification is
considered as the most fundamental theoretical basis. With this
consideration as a starting point, we have found that the one-loop
RG study, involving SUSY $SU(5)$ degrees of freedom placed near
$M_{GUT}$, allows one to select some specific classes of the scale
hierarchies. The proton stability is provided simultaneously.

It was shown that states near $M_{GUT}$ should be taken
into account as threshold corrections; they are crucially
important for the selection of scales, giving sufficiently high
unification point. Then the RG consideration at the one-loop level
results in two classes of scenarios with an opposite arrangement of
$\mu$ and $M_{1/2}$ scales. At the same time, the analysis does not
fix the characteristic scalar scale $M_0$ due to a specific form of
the RG equations which contain the squark and slepton scales as a
ratio $M_{\tilde q}/M_{\tilde l}$ only.

For the first class of scenarios the hierarchy $|\mu|\gg M_{1/2}$
takes place. The second class is defined by the hierarchy $|\mu|\ll
M_{1/2}$. Due to arbitrariness of $M_0$ it is possible to select
some subscenarios with various $M_0$ arrangements. Among them there
can be found some hierarchies corresponding to Split Symmetry and
some ideologically close scenarios which do not reject fine-tuning
and are motivated, on the one hand, by the anthropic principle and,
on the other hand, multi-vacua string landscape
arguments~\cite{1,2,3,4,5,6} (see, however, comments in~\cite{7}).

In particular, the RG analysis results in the hierarchy
\begin{eqnarray}
 \displaystyle M_{0}\sim
   M_{1/2}\gg |\mu| > M_{EW},
 \label{M}
\end{eqnarray}
whose spectrum contains two lightest neutralinos degenerated in mass
(almost pure Higgsino) and one of charginos as the states that are the nearest
to the electroweak scale.
In this Split Higgsino scenario both $M_0$ and
$M_{1/2}$ are shifted to scales $\sim (10^7 - 10^{10}) \, \mathrm{TeV}$.

In this paper, we consider some features and manifestations of
the last scenario only (a preliminary version of this work was presented
in~\cite{8}, see also~\cite{9}).

As it will be shown, direct observation of neutralino (Higgsino) in
$\chi-N$ scattering is impossible nowadays due to a very small
interaction cross section. As to collider experiments, the signature
of neutralino and chargino production and decays at the LHC
crucially depends on the mass splitting values for these degenerate
states~\cite{9,10,11,12,13}. On the one hand, observation of these
hardly detectable effects means that it is possible to get important
information on the high superstate structure, in particular, $\tilde
t_{1,2}$ and their mixing, from one-loop calculations of the mass
splitting~\cite{14,15,16}. On the other hand, if only these
low-lying SUSY degrees of freedom are detected in experiments near a
TeV scale, the conventional MSSM spectrum should necessarily be
splitted in some manner to produce a high scale for other
superstates. The value predicted for the diffuse gamma flux from the
Galactic halo can be measured by GLAST or ACT in the near future but
it is not peculiar to this scenario. Signals from direct photons do
not contain any specific effects too ~\cite{9}.

Thus, the Split Higgsino scenario is an example of SUSY models, in
which the lightest degrees of freedom manifest themselves in some
subtle effects as specific decay channels, correlating with the
value of diffuse (or direct) gamma flux. Note, even if collider
experiments will demonstrate the absence of the lowest neutralino
signals (due to a high degeneracy of states, for example, it will
certainly testify to some fine-tuning in the SUSY states spectrum),
the SUSY, as a theoretical principle, cannot be rejected. At low
$O\,(1\,\mathrm{TeV})$ energies the SUSY, in a sense, is barely
hidden, so it exists at a high scale.

The structure of the paper is as follows. In Section 2, we discuss
RG analysis of the SUSY $SU(5)$ in detail, indicating the most
interesting scenarios. In Section 3, the Lagrangian of the Split
Higgsino model and its analysis are presented. There is also an
estimation of the lowest Higgsino mass from the relic abundance
data. We discuss expected experimental manifestations of the
scenario in Section 4 and some conclusions are in Section 5.

\section{Renormgroup analysis}

We consider one-loop RG equations for running couplings and start from their values
at the $M_Z$ scale
\begin{eqnarray}
 \displaystyle \alpha^{-1}(M_Z)=127.922\pm 0.027,
  & & \quad \alpha_s(M_Z)=0.1200\pm0.0028, \nonumber \\
      \quad \sin^2\theta_W(M_Z)
  &=& 0.23113 \pm 0.00015.
\end{eqnarray}
In renormgroup equations the following values are used as initial:
\begin{eqnarray}  \vspace*{5mm}
  \displaystyle \alpha_1^{-1}(M_Z)=\frac35\alpha^{-1}(M_Z)\cos^2\theta_W(M_Z)
  &=& 59.0132 \mp(0.0384)_{\sin^2\theta_W}\pm(0.0124)_\alpha, \nonumber
\\ \vspace*{5mm}
  \displaystyle \alpha_2^{-1}(M_Z)=\alpha^{-1}(M_Z)\sin^2\theta_W(M_Z)
  &=& 29.5666\pm(0.0192)_{\sin^2\theta_W}\pm0.0062)_\alpha, \label{exp}
\\
  \alpha_3^{-1}(M_Z)=\alpha_s^{-1}(M_Z)
  &=& 8.3333\pm0.1944.
  \nonumber
\end{eqnarray}
At the one-loop level the equations for running couplings are well-known
\begin{eqnarray} \displaystyle
\alpha_i^{-1}(Q_2)=\alpha_i^{-1}(Q_1)+\frac{b_i}{2\pi}\ln\frac{Q_2}{Q_1},\qquad
b_i=\sum_jb_{ij}. \label{b} \end{eqnarray} In the sum all states
with masses $M_j<Q_2/2$ at $Q_2>Q_1$ are taken into account.

Namely, we include in the RG analysis the following
degrees of freedom: singlet quarks and their superpartners
$(D_L,\,\tilde D_L),\;(D_R,\,\tilde D_R)$ contained in super-Higgs
quintets of SUSY $SU(5)$; chiral superfields $(
\Phi_L,\,\tilde\Phi_L)$ and $(\Psi_L,\,\tilde\Psi_L)$ in adjoint
representations of $SU(2)$ and $SU(3)$, respectively, which survive
from super Higgs 24-plet. In the minimal SUSY $SU(5)$ masses
of the states $M_5=(M_D,\,M_{\tilde D}),\;
M_{24}=(M_{\Psi},\,M_{\tilde\Psi},\,M_{\Phi},\, M_{\tilde \Phi})$
are generated by an interaction with Higgs condensate at the GUT scale,
but the interaction couplings are not fixed. For the analysis we
assume masses $M_5,\,M_{24}$ to be slightly lower than $ M_{GUT}$.
Note that heavy states near $M_{GUT}$ are of especial
importance for the RG equations. The necessity of taking into account of such
heavy states in the RG analysis as so-called threshold corrections
is well-known~\cite{17,18,19,20,21,22}.

For one-loop running couplings  at the scale $q^2=(2M_{GUT})^2$
we have
\begin{eqnarray}
 \begin{array}{c}  \vspace*{5mm}
    \displaystyle \alpha_1^{-1}(2M_{GUT})=\alpha_1^{-1}(M_Z)-\frac{103}{60\pi}\ln 2+
    \frac{1}{2\pi}\left(-7\ln M_{GUT}+\frac{4}{15}\ln
    M_{D}+\frac{2}{15}\ln M_{\tilde D}\right.
\\ \vspace*{5mm}
    \displaystyle \left. +\frac{11}{10}\ln M_{\tilde q} +\frac{9}{10}\ln
    M_{\tilde l}+\frac25\ln \mu +\frac{1}{10}\ln M_{H} +\frac{17}{30}\ln
    M_t+\frac{53}{15}\ln M_Z \right),
\\ \vspace*{5mm}
    \displaystyle \alpha_2^{-1}(2M_{GUT})=\alpha_2^{-1}(M_Z)-\frac{7}{4\pi}\ln 2+
    \frac{1}{2\pi}\left(-3\ln M_{GUT}+\frac43\ln
    M_{\tilde\Phi}+\frac23\ln M_{\Phi}\right.
\\ \vspace*{5mm}
    \displaystyle \left.+\frac{3}{2}\ln M_{\tilde q} +\frac{1}{2}\ln M_{\tilde
    l}+\frac43\ln M_{\tilde W}+\frac23\ln \mu +\frac{1}{6}\ln M_{H}
    +\frac{1}{2}\ln M_t-\frac{11}{3}\ln M_Z \right),
\\ \vspace*{5mm}
    \displaystyle \alpha_3^{-1}(2M_{GUT})=\alpha_3^{-1}(M_Z)+\frac{23}{6\pi}\ln 2+
    \frac{1}{2\pi}\left(-\ln M_{GUT}+2\ln M_{\tilde\Psi}+\ln M_\Psi \right.
\\
    \displaystyle \left.+\frac{2}{3}\ln M_{D}+\frac{1}{3}\ln M_{\tilde D}+ 2\ln
    M_{\tilde q} +2\ln M_{\tilde g}+\frac23\ln M_t-\frac{23}{3}\ln M_Z \right).
 \end{array}
 \label{gut}
\end{eqnarray}
Here $M_0=(M_{\tilde q},\,M_{\tilde l})$ are masses of scalar quarks
and leptons averaged over chiralities and generations; $M_t$ is
$t$-quark mass; other parameters were introduced above. In
(\ref{gut}) it is supposed that the lightest Higgs boson mass $m_h$
is near to $M_{EW}$ and other Higgs bosons $H,\,A,\,H^\pm$ are
placed at some high $M_{H}$ scale. For extra heavy states the only
condition is $M_{24},\,\,M_5 < M_{GUT}$ and we suppose that this
inequality is fulfilled with accuracy within 1 - 2 orders of
magnitude. As to SUSY degrees of freedom
$M_0,\,M_{1/2},\,\mu,\,M_{H}$, equations (\ref{gut}) do not depend
on any specific arrangement of these scales. So when $M_{GUT}$ is
the highest scale in the system, there is no necessity to establish
an initial scale hierarchy. The only demand is that the RG equations
should lead to coupling unification at sufficiently high $M_{GUT}$
scale for the proton stability. Then the set of possible hierarchies
of energy scales arises as the final result of the study. Note also
that if masses of singlet superstates and residual Higgs superfields
are equal to $M_{GUT}$, equations (\ref{gut}) return to the standard
form automatically.

The above considerations together with experimental restrictions on
the SUSY mass spectrum and general expectation of the lightest Higgs
boson scale (it is not very far from the EW scale) are sufficient
for the RG analysis.

As the first step, all couplings were recalculated at the $2M_Z$
scale, all the SM states contribute to running of couplings despite
$W^\pm,\,Z^0$, Higgs bosons, and $t$-quark. At the same time, terms
with $\ln 2$ occur which are quantitatively important for
calculations. Between the $(2M_Z,\;2M_t)$ scales the following
states emerge: $W^\pm,\,Z^0$ and one Higgs doublet containing light
$h$-boson and longitudinal degrees of freedom of $W^\pm,\,Z^0$. At
these stages, $Z\bar qq$ vertex was used for calculations of
$\alpha_2^{-1}(2M_t)$, starting from $\alpha_2^{-1}(2M_Z)$ value.
Above $2M_t$ calculations were carried out in a standard manner.

Now, equating couplings at $M_{GUT}$, from (\ref{gut}) we get the
following expressions:
\begin{eqnarray}
\displaystyle
M_{GUT}=Ak_1M_Z\left(\frac{M_Z}{M'_{1/2}}\right)^{2/9},\qquad
\mu=Bk_2M_Z\left(\frac{M_Z}{M'_{1/2}}\right)^{1/3}, \label{3}
\end{eqnarray}
where
\begin{eqnarray} \vspace*{5mm}
 \displaystyle k_1
    &=& K_{\tilde q\tilde l}^{-1/12}K_{GUT1}^{1/3}\equiv
        \left(\frac{M_{\tilde l }}{M_{\tilde
        q}}\right)^{1/12}\left(\frac{M_{GUT}}{M'_{GUT}}\right)^{1/3}, \nonumber
\\
 \displaystyle k_2
    &=& K_{Ht}^{-1/4}K_{\tilde q\tilde l}^{1/4}K_{\tilde g\tilde
        W}^{5/2}K_{GUT2}^{-1}\equiv \left(\frac{M_{top}}{M_{H}}\right)^{1/4}
        \left(\frac{M_{\tilde q }}{M_{\tilde l}}\right)^{1/4}
        \left(\frac{M_{\tilde g }}{M_{\tilde W}}\right)^{5/2}
        \left(\frac{M''_{GUT}}{M_{GUT}}\right), \nonumber
 \label{4}
\end{eqnarray}

\[
 \begin{array}{c} \vspace*{4mm}
  \displaystyle M'_{1/2}\equiv (M_{\tilde W}M_{\tilde g})^{1/2},\qquad M'_{GUT} \equiv
  (M_{\tilde \Psi}M_{\tilde \Phi})^{1/3}(M_{\Psi}M_{\Phi})^{1/6}\leq M_{GUT},
\\ \vspace*{2mm}
  \displaystyle M''_{GUT} \equiv \frac{(M^2_{\tilde
  \Psi}M_{\Psi})^{7/6}(M_D^2M_{\tilde D})^{1/2}}{(M^2_{\tilde
  \Phi}M_{\Phi})^{4/3}}\leq M_{GUT},
\\ \vspace*{3mm}
  \displaystyle A=\exp\left(\frac{\pi}{18}(5\alpha_1^{-1}(M_Z)-3\alpha_2^{-1}(M_Z)-
  2\alpha_3^{-1}(M_Z))-\frac{11}{18}\ln2\right)=(1.57\times ^{1.09}_{0.92})\cdot 10^{14},
\\
  \displaystyle B=\exp\left(\frac{\pi}{3}(5\alpha_1^{-1}(M_Z)-12\alpha_2^{-1}(M_Z)+
  7\alpha_3^{-1}(M_Z))+\frac{157}{12}\ln2\right)=(2.0\times ^{0.15}_{6.56})\cdot 10^{3}.
 \end{array}
 \label{5}
\]

Here all dimensionless parameters $K$ with various indices are
defined as quantities having values larger than unity (see
also~\cite{23}). Note that $K_{GUT1},\,K_{GUT2}$ are not under
theoretical control either in the MSSM or in the SUSY $SU(5)$. So we
assume that $1\le K_{GUT1}\simeq K_{GUT2}\le 10$. For the lightest
Higgs boson there is an experimental restriction \cite{24}:
$M_{h}>114.4 \;\mathrm{GeV}$, as to other (heavy) Higgs bosons the
following interval $2\le K_{Ht}\le 10$ is supposed for the numerical
analysis. Values of $K_{\tilde q\tilde l},\,K_{\tilde g\tilde W}$
can be determined from the renormgroup evolution from $M_{GUT}$ to
$M_0,\,M_{1/2}$. Here we suppose that $1.5\le K_{\tilde q\tilde
l}\simeq K_{\tilde g\tilde W}\le 2.5$. Now all dimensionless
parameters are fixed in some intervals and we analyze $M'_{1/2}$ and
$\mu$ as functions of $M_{GUT}$:
\begin{eqnarray}
 \displaystyle M'_{1/2}(M_{GUT})=(Ak_1)^{9/2}M_Z^{11/2}\times
 M_{GUT}^{-9/2},\qquad \mu(M_{GUT})=
 \frac{Bk_2}{(Ak_1)^{3/2}M_Z^{1/2}}\times M_{GUT}^{3/2}.
 \label{6}
\end{eqnarray}

As it is seen, the initial system of equations for three running couplings can be
rewritten as a system of two equations for the effective gaugino
$M'_{1/2}$ and $\mu$ scales, depending on $M_{GUT}$. It is very
essential that all other characteristic scales arise in the
equations as dimensionless ratios only. (This method of RG analysis
was used in~\cite{23} for investigation of SUSY $SU(5)$ and
$E_6$ energy scale hierarchies; see also~\cite{20,22}.)

As an important feature, we have to note that the renormgroup study
remains a common scalar scale $M_0$ as an arbitrary one: the scale
$M_0$ turns into (\ref{6}) through the ratio $M_{\tilde q}/M_{\tilde
l}$ only. For numerical predictions the ratio value was estimated as
$O(1)$. Certainly, the splitting of the $M_{\tilde q}$ and
$M_{\tilde l}$ scales will affect the hierarchies of other scales.
However, to change these hierarchies crucially, the ratio $M_{\tilde
q}/M_{\tilde l}$ must be $\sim O(100)$ or larger, and here we do not
consider this possibility. Also, we suppose that radiative
corrections to superscalar masses do not change the ratio
substantially (numerically, these corrections can be as large as
$(1-5)\,\%$, see~\cite{14}). Conclude, the scale $M_0$ should be
fixed by some extra arguments independently.

Contributions of heavy states $M_{24}, \, M_5$ do not affect the RG
hierarchy of SUSY scales radically because their scales are
combined into a ratio too. Nevertheless, the ratio value is
important to fix the $M_{GUT}$ scale: the ratio $M_{24}/M_5$ depends on
$M_Z$ logarithmically, so numerical coefficients of the ratio can
change the $M_{GUT}$ value.

For the analysis we used the known restrictions for the proton lifetime
$(\tau_p\geq 10^{32}\;\mathrm{yr}$ at $M_{GUT}\geq
10^{15}\;\mathrm{GeV})$ and for $M_{SUSY}$ that is $\sim M'_{1/2}>100
\;\mathrm{GeV}$ when $M_{GUT}< 3\cdot 10^{16} \;\mathrm{GeV})$. We
hope also that loop corrections to all masses in the model do not
change results of the RG analysis drastically, it is supposed that for
the scenario they contribute no more than $\sim (10-15)\,\%$~\cite{14,15}.
\begin{figure}[h]
\begin{minipage}{1.3\textwidth}
\epsfxsize=\textwidth\epsfbox{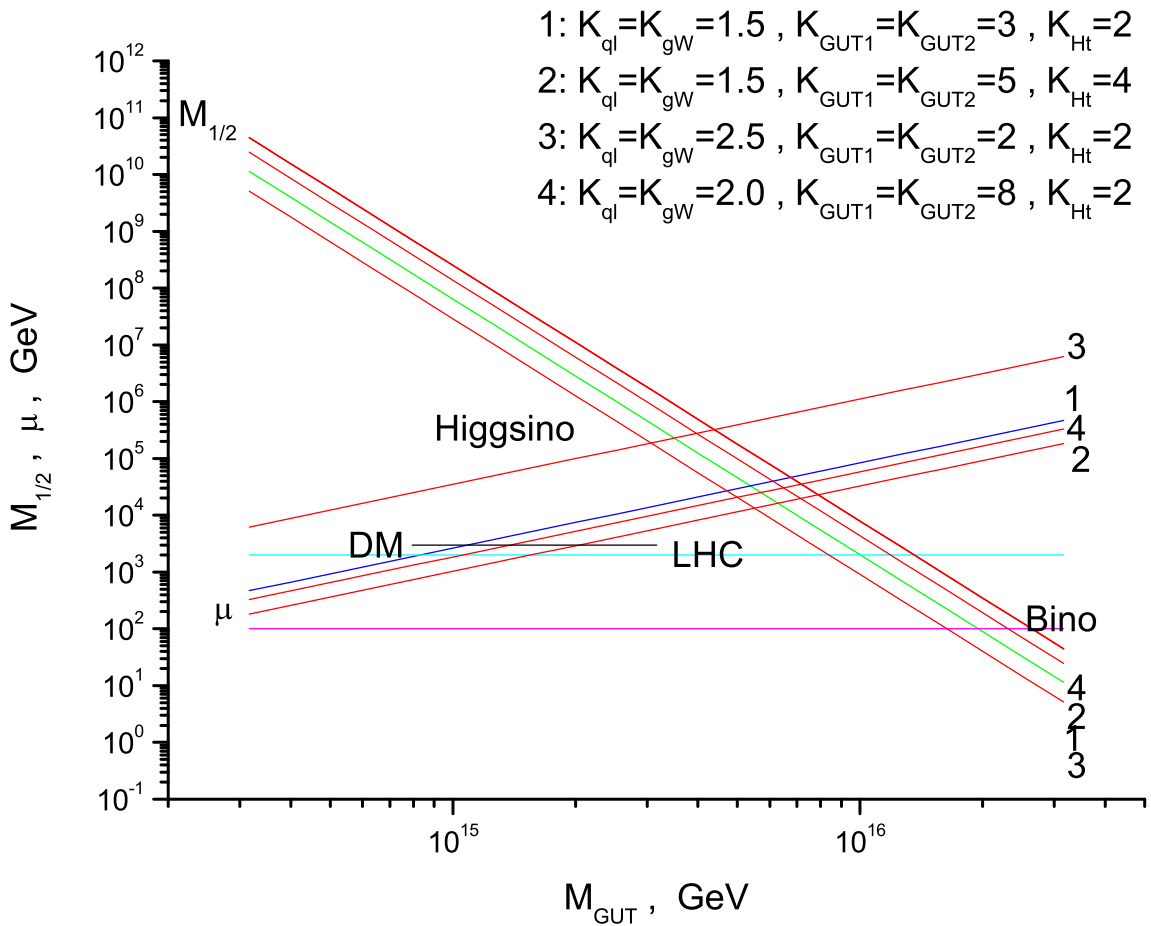}
\end{minipage}
   \caption{\label{fig:fig2} \small SUSY $SU(5)$ hierarchies of scales
   permitted by the one-loop RG analysis.}
\end{figure}

Numerical results following from (\ref{6}) are shown in
Fig.~\ref{fig:fig2}. It is clearly seen that the proposed method of
the one-loop RG analysis results in the selection of two proper sets
of SUSY scales only (further we will omit differences between
$M'_{1/2}$ and $M_{1/2}$). For this conclusion we assume that the
characteristic $\mu$ scale is in the $O(1\,\mathrm{TeV})$ region,
providing the neutralino mass value that is sufficient for
explanation of the DM properties.

The first class of scenarios is defined by the hierarchy $$|\mu| \gg
M_{1/2} \gtrsim M_{EW}.$$ As it was noted, $M_0$ value does not
affect this scale splitting dictated by the one-loop RG evolution,
and can be placed at an arbitrary but physically reasonable scale.
So the class could be divided into some subscenarios with
hierarchies
\begin{eqnarray*}
&&(a)\quad M_0\gg |\mu| \gg
M_{1/2} \gtrsim M_{EW},\\
&&(b)\quad M_0\sim |\mu| \gg
M_{1/2} \gtrsim M_{EW},\\
&&(c)\quad |\mu| \gg M_0\gg
M_{1/2} \gtrsim M_{EW},\\
&&(d)\quad |\mu| \gg
M_0\sim M_{1/2} \gtrsim M_{EW}.
\end{eqnarray*}
The subscenarios (a), (b) and (c) are close to the known Split SUSY
(Gaugino) models~\cite{1,2,3}, but the variant (d), keeping most
part of the SUSY degrees of freedom (except Higgsino) near the EW scale,
is similar to the MSSM spectrum with ``naturalness''. For these
variants we get $M_{GUT}=(1-5)\cdot 10^{16}\,\mathrm{GeV}$ and
gaugino masses can be as small as $(0.1-1.0)\,\mathrm{TeV}$.

The second set of scenarios is generated by the hierarchy
$$M_{1/2}\gg |\mu| > M_{EW}.$$ In this set we have the following
subscenarios:
\begin{eqnarray*}
&&(e)\quad M_0\gtrsim M_{1/2}\gg |\mu|> M_{EW},\\
&&(f)\quad M_{1/2}\gg M_0 \sim |\mu| > M_{EW},\\
&&(g)\quad M_{1/2}\gg M_0 \gg|\mu| > M_{EW},\\
&&(h)\quad M_{1/2}\gg |\mu| >M_0\gtrsim M_{EW}.
\end{eqnarray*}

Except the last variant (h) with too low $M_0$ scale as
contradicting known experimental data, the other scenarios of this
class seem reasonable. All of them introduce Higgsino as the DM
carrier. Note that the hierarchy (f) is close to the MSSM spectrum
again, despite gaugino states at a high scale. Scenarios (e) and (g)
shift $M_0$ and $M_{1/2}$ scales into a multi-TeV region, splitting
the spectrum. In these cases, the $\mu$ scale is the nearest to the
EW scale and should be near $1\,\mathrm{TeV}$, as it follows from
the RG analysis. As to the unification scale, we have the value
$M_{GUT} = (1-2)\cdot 10^{15}\,\mathrm{GeV}$ which makes the proton
stable.

Obviously, Split SUSY models with high $M_0$ and $M_{1/2}$ scales
lead to the damping of heavy states (squarks, sleptons)
contributions near the EW scale. Another important consequence of
this type of scenarios is (nearly) degeneration in mass of the
lowest neutralino and chargino states ~\cite{9,11,13,16,25,26}
(experimental searches of this degenerated states were carried out
at LEP energies~\cite{27} without any evident signals).

Further, we will consider the Split Higgsino scenario with hierarchy
$$M_{0}\gtrsim M_{1/2}\gg |\mu|>M_{EW}.$$

Taking $M_{GUT}$ value as $\gtrsim 10^{15}\;\mathrm {GeV}$, for $\mu
=1.0 \;\mathrm{TeV}$ the SUSY breaking scale is $M_{SUSY} \sim
M_{1/2}\sim (0.1 - 2.5)\cdot 10^8\;\mathrm{GeV}$; when $\mu =1.4
\;\mathrm {TeV}$ we get $M_{SUSY}\sim (0.5 - 2.8)\cdot
10^8\;\mathrm{GeV}$. For these cases $M_{GUT} \sim(1.0 - 1.7)\cdot
10^{15}\;\mathrm {GeV}$. The used values of $\mu$, as it will be
seen further, agree with modern data on relic DM abundance.
Intervals of $M_{SUSY}$ values follow mainly from uncertainties in
the parameters $K_{GUT1}$ and $K_{GUT2}$. Most evidently, the
scenario is selected when $K_{GUT1},\,K_{GUT2}\sim 5-8$; it means
that the effect of threshold corrections is essentially important
for the discovery of hierarchy. Namely, if masses $M_{24}, \, M_5$
are equal to $M_{GUT}$, it is questionable to select several
possible scenarios that are compatible with the DM data.

These one-loop results should be refined with taking into account of
two-loop $\beta$-functions and loop corrections to masses. By now we
expect that two-loop RG analysis does not change the above
hierarchies radically as well as mass corrections. Certainly,
various numerical coefficients will change, arrangements of the
scales will move somehow, but the relative spacing between $\mu$ and
$M_{1/2}$ scales will not change, as we hope. Nevertheless, the
$M_0$ scale can be splitted by the RG at the two-loop level, so
positions of the $M_{\tilde q}$ and $M_{\tilde l}$ scales can be
determined independently. More carefully this question will be
analyzed in a forthcoming paper. Now we will consider the Lagrangian
of the Split Higgsino scenario, its structure, and features.

\section{The Split Higgsino scenario}
\subsection{The model Lagrangian and its features}

When the group of symmetry is fixed, the Lagrangian as a local gauge
group invariant can be written in a standard manner. Nevertheless,
its specific form depends on the chosen field representation. In
this paper, we use the formalism of Majorana spinors and define all
physical fields as having positive masses only.

More precisely, in the scenario considered the neutralino mixing
parameter is proportional to the ratio $M_Z/M_{SUSY}$, so
it is negligibly small due to a high scale of SUSY breaking.
However, the mixing can be important when the model restrictions are
analyzed in the framework of a specific mass spectrum with a small
splitting of the lightest neutralino masses.

As to neutralino masses, i.e., the scale of $\mu$ parameter,
diagonalization of the neutralino mass matrix by an orthogonal
matrix usually leads to emerging of neutralino state with negative
mass. This should be taken into account in calculations by the
corresponding definition of the propagator (if negative mass
neutralino is in the intermediate state) or in the neutralino
polarization matrix (if the neutralino is in the initial or in the
final state). To evade this inconvenience, we redefine the
neutralino field with the negative mass in the following way: $\chi
\rightarrow i\gamma_5\chi$. This operation provides positive
neutralino mass keeping standard calculation rules simultaneously.
Moreover, it does not change Majorana type of the field. It also has
been shown that this procedure is equivalent to the diagonalization
by the unitary matrix instead of the orthogonal one, for details see
~\cite{28}.

Hence, in the scenario two lowest neutralino states are Majorana
Higgsino-like ones. The set of scales leads to the strong splitting
in the neutralino and chargino spectra, so masses of these lightest
neutralinos $\chi^0_{1,\,2}$ and the light chargino $\chi^\pm_1$
emerge near the $\mu$ scale. Light neutralino states are built from
the initial fields $h_{1,2}$ in the limit of pure Higgsino when
$m_Z/\mu\rightarrow 0$ and $m_Z/M_{SUSY}\rightarrow 0$; here
$M_{SUSY} \sim M_1 \sim M_2$. Masses of the lightest states are
\begin{eqnarray}  \vspace*{5mm}
 \displaystyle M_{\chi^0_1}
  &\simeq& \mu -\frac{M_Z^2(1+\sin
           2\beta)}{2M_1M_2}(M_1\cos^2\theta_W+M_2\sin^2\theta_W), \nonumber
\\ \vspace*{5mm}
 \displaystyle M_{\chi^0_2}
  &\simeq& \mu+\frac{M_Z^2(1-\sin
           2\beta)}{2M_1M_2}(M_1\cos^2\theta_W+M_2\sin^2\theta_W),
\label{H}
\\
 \displaystyle M_{\chi^\pm_1}
  &\simeq& \mu -\mu \frac{M_W^2}{M_2^2}-M_W \frac{M_W}{M_2}\sin2\beta .
\nonumber
\end{eqnarray}

It is seen that the spectrum of the lowest states, $\chi^0_{1,\,2}$
and $\chi^\pm_1$, is nearly degenerated. For the Higgsino DM this
fact is known (see, for example,~\cite{16,25,26,29}). Two other
neutralino states, $\chi^0_{3,\,4}$, and heavy chargino $\chi^\pm_2$
are placed far from the lightest ones, near the high scale $M_{SUSY}$.

The initial Lagrangian of $Z-h_{1,2}$ interactions, including the
$\mu$-term, is the following (for more details see~\cite{28})
\begin{equation}
 L=\frac{1}{2}g_Z Z_{\mu}(\bar{h}_{1L}\gamma^{\mu}h_{1L}+
 \bar{h}_{2R}\gamma^{\mu}h_{2R})-
  \mu(\bar{h}_{1R}h_{2L}+\bar{h}_{1L}h_{2R}).
  \label{L1}
\end{equation}
Taking into account that $\bar h_{aL}\gamma^{\mu}h_{aL}= -\bar
h_{aR}\gamma^{\mu}h_{aR},\,\bar h_{1R}h_{2L}=\bar h_{2R}h_{1L},$
from (\ref{L1}) in the limit of pure neutral Higgsino it follows:
\begin{equation}
 L=-\frac{1}{2}g_Z Z_{\mu}\bar{h}_D\gamma^{\mu}h_D-\mu\bar{h}_Dh_D,\label{L2}
\end{equation}
where $h_D=h_{1R}+h_{2L}$ is the Dirac neutral field and $g_Z=g_2/\cos
\theta_W$. When the field $h_D$ is expressed through Majorana fields
$\chi_1$ and $\chi_2$ as
\begin{equation}
 h_D\equiv h_{1R}+h_{2L}=(\chi_1-i\chi_2)/\sqrt{2},
\end{equation}
for the transition $(h_1, h_2)\rightarrow (\chi_1, \chi_2)$ we get
\begin{equation}
 h_{1,2}=\frac{1}{\sqrt{2}}(\chi_1\mp i\gamma_5\chi_2);\qquad
 \chi_1=\frac{1}{\sqrt{2}}(h_1+h_2),\quad\chi_2=\frac{i}{\sqrt{2}}(h_1-h_2).
\end{equation}

All processes near the EW scale are described by the SM Lagrangian
together with extra Lagrangian of the Higgsino interactions with
photons and vector bosons:
\begin{eqnarray}  \vspace*{5mm}
 \displaystyle \Delta L
  &\simeq& -\left(eA_\mu+\frac{g_2}{2\cos\theta}\cos2\theta Z_\mu
      \right) \bar \chi_c\ga^\mu\chi_c +
      \frac{ig_2}{2\cos\theta}Z_\mu\bar \chi^0_{1}\ga^\mu \chi^0_{2}+
      \nonumber
\\
 \displaystyle
  &-& \frac{g_2}{2}W^+_\mu(i\bar \chi^0_1+\bar
      \chi^0_2)\ga^\mu \chi_c+\frac{g_2}{2}W^-_\mu \bar
      \chi_c \ga^\mu (i\chi^0_1-\chi^0_2),
 \label{LH}
\end{eqnarray}
where $\chi_c$ denotes the lowest chargino state, $\chi^{\pm}_1$.
Remember that $\chi^0_{1,\,2}$ states are the Majorana spinors and
chargino $\chi_c$ is the Dirac spinor.

In the above Lagrangian the vertices are considered in the limit of
zero mixing, because contributions to the vertices induced by a
mixing are of an order of $M_Z/M_{SUSY}$. So corrections induced by
mixing are negligibly small and can be omitted. However, the same
small mixing contributions to the particle mass spectrum should be
involved into analysis. The reason is that the decay widths depend
on the mass splitting crucially.

\subsection{Split Higgsino as the DM carrier}

Supposing that the lightest Higgsino-like neutralino is a carrier
of the DM in the Universe, we evaluate its mass from the DM relic
abundance.

In accordance with the known method for the relic abundance
calculation (see, for example,~\cite{30,31,32,33} and~\cite{34} with
references therein) before freeze-out neutralinos are in the
thermodynamical equilibrium with other components of the
cosmological plasma. Relic is formed by the irreversible
annihilation process -- from the moment, when the freeze-out is
reached, up to the present day when temperature is approximately
equal to absolute zero (this approximation is sufficient for the
case). With the standard estimation of $x_f = M_{\chi}/T_f \approx
20 - 25$ the value of $T_f $ can be well above $T_{EW}\sim 100
\;\mathrm{GeV}$ only if $m_{\chi} > 2\;\mathrm{TeV}$. Now, if the
lowest Higgsino masses were well above $T_{EW}$, the irreversible
annihilation process could start before electro-weak phase
transition (it is considered as the first order phase transition) in
the high-symmetric phase of the cosmological plasma. Then the plasma
does not contain any Higgs condensate, and all standard particles,
except Higgsino, are massless (more exactly, their masses $\ll T$).

In the ordinary low-symmetric case, the neutralino annihilation
cross section is calculated with the Lagrangian (\ref{LH}).
In relic calculations all possible coannihilation channels
were taken into account (there is no coannihilation with squarks and/or
sleptons), namely
\begin{equation}
\begin{array}{l}
\displaystyle \chi_{\alpha} \bar \chi_{\beta} \rightarrow ZZ, \quad
W^+W^-
\\[3mm]
\displaystyle \chi _c\bar \chi_c \,\,\rightarrow ZZ,\quad W^+W^-,
\quad f \bar f, \quad \gamma \gamma,\quad Zh
\\[3mm]
\displaystyle \chi_{\alpha} \bar \chi_c \,\,\rightarrow ZW, \quad l
\nu_l, \quad q_i \bar q_j, \quad\gamma W,\quad Wh,
\end{array}
\label{coann}
\end{equation}
where $\alpha,\, \beta\, = 1,\, 2$ denote the lowest neutralino states.

In the scenario the effective neutralino annihilation cross section
is:
\begin{equation}
\begin{array}{c}
\displaystyle <(\si v)_{ann}> = \frac{g_2^4}{128\; \pi M_{\chi}^2}\cdot \left\{27+
4(3+5t_W^4)(c_W^4+s_W^4) - kc_W^2+\frac{k^2}{4}(c_W^4+s_W^4)+\right.
\\[5mm]
\displaystyle \left.
\frac{1}{2\ c_W^4}\left[1+\frac{1}{8}(c_W^4+s_W^4) +
(c_W^2 - s_W^2)^4\right]+\frac{1}{2\; c_W^2}\left[s_W^4\left(10 - \frac{1}{k}\right)
+ 2kc_W^2\right]\right\},
\end{array}
\label{sLS}
\end{equation}
where $t_W = \tan \Theta_W, \,s_W = \sin \Theta_W,\, c_W = \cos
\Theta_W$ and $k = M_Z^2/M_W^2$. To compare the calculated value of
$\Omega h^2$ with an experimental corridor of the relic data
~\cite{35}, we use known values of the above parameters and extract
the following LSP (Higgsino) mass: $M_{\chi} = 1.0 - 1.4
\;\mathrm{TeV}$ for $x_f=25$ and $M_{\chi} = 1.4 - 1.6
\;\mathrm{TeV}$ for $x_f=20$. These values do not spoil the gauge
coupling convergence and are in agreement with the results from
~\cite{2,3,9,29}. Thus, in the model where two lightest neutralinos
and one chargino are the closest to the EW scale having masses
$O(1\;\mathrm{TeV})$. Further, we will use $M_{\chi} = 1.4
\;\mathrm{TeV}$ for all numerical estimations as an average value.

To answer whether the neutralino annihilation process can start
in the high-symmetric phase, we calculate the annihilation cross section with the
Lagrangian which contains physical states presented by chiral
fermions and gauge fields $B,W_a \,(a=1,2,3)$. Due to the absence
of the Higgs condensate neutralino and chargino degrees of freedom
join into the fundamental $SU(2)$ representation, i.e., Dirac field
$\chi_D$. All states of this field are dynamically equivalent and
correspond to quantum numbers of the restored (unbroken) $SU(2)$ symmetry.

In t- and s-channels all cross sections of Higgsino annihilation
into gauge bosons and massless fermions were calculated analogously
to the QCD calculation technology. The only difference is that it is
necessary to consider all channels with initial and final states
which have an arbitrary color in two dimensions, corresponding to
the restored $SU(2)$ symmetry. Calculating the total neutralino
annihilation cross section, from the known corridor of the relic
abundance values we get the lowest Higgsino mass $\sim
1\;\mathrm{TeV}$ again. It contradicts the initial supposition that
$M_{\chi}> 2 \;\mathrm{TeV}$ to provide freeze-out temperature
higher than the EW phase transition temperature. So in the framework
of orthodox notions (mechanisms of extra entropy production or
superheavy states decays, etc. are not considered) this scenario
does not allow the DM to be created in the high symmetric phase.

\section {SUSY scales and experimental possibilities}

\subsection{$\chi-N$ scattering and collider signals}

It is important to understand how an experimental study of the
Split Higgsino scenario can be realized. Here we discuss some
possibilities that can be given by neutralino-nucleon scattering
processes, collider experiments with SUSY particles creation
and the data on photon spectrum from neutralino annihilation.
The last one will be considered in the next subsection.

An experimental observation of the scenario manifestations depends
on neutralino mass splitting parameters $\Delta
M_{\chi^0}=M_{\chi^0_2} - M_{\chi^0_1}$ and $\Delta M_{\chi_c} =
M_{\chi_c}- M_{\chi^0_2}$ crucially. These splittings are determined
by the sum of tree splittings and radiative corrections to masses.
Depending on the $M_0$ spacing and structure of high energy states,
loop corrections can be comparable with tree values of $\Delta M_{\chi}$
($\Delta M_{\chi^0}$ or $\Delta M_{\chi_c}$) or even exceed
them~\cite{16,12,14,15,26}. As it is known~\cite{16}, the hierarchy of
$\tilde t_1$ and $\tilde t_2$ states and their mixing angle
$\Theta_t$ drive the value of the mass difference when squarks dominate in
loops
$\Delta M_{\chi_c}\sim \Delta M_{\chi^0}\sim
m_t^3\sin(2\Theta_t)\cdot \ln(m^2_{\tilde{t}_1}/m^2_{\tilde{t}_2}).$
Due to these (large) corrections the mass splittings can
be increased up to $\sim 10\, \mathrm{GeV}$~\cite{16,26}. Then
let us consider two possible variants.

Firstly, let there be large mass splittings $\Delta M_{\chi}\sim (1
- 10) \; \mathrm{GeV}$ or even larger. When this is the case, the
lowest Higgsino and chargino states can be detected at the LHC, in
particular, due to specific decay channels of $\chi^0_2$ and
$\chi_c$ -- the corresponding analysis was considered in some detail
in~\cite{9,10,11,12}, so we do not repeat it here. In the Split
Higgsino hierarchy $M_0\gtrsim M_{1/2}\gg \mu$, squarks (sleptons)
slip out of the LHC experiment (together with the high energy
gaugino), keeping specific chargino (neutralino) decays as the
observable only.

If, however, $M_{1/2}\gg M_0$, there are two additional
possibilities. Namely, if $M_{1/2}\gg M_0 \sim \mu$, i.e., squarks
(sleptons) have masses $\sim (1-2) \;\mathrm{TeV}$, an occasion
arises to observe their signals at the LHC. (If $m_{\bar q}$ and/or
$m_{\bar l}$ are sufficiently close to the lowest neutralino mass,
some coannihilation contribution to the effective neutralino
annihilation cross section is produced.) When $M_{1/2}\gg M_0\gg\mu$
there are no observable effects of squarks (sleptons) at the LHC
scale.

Both of these last subscenarios are very peculiar due to a large
splitting between $M_{1/2}$ and $M_0$ -- it can produce relatively
long-lived superscalars at TeV (or higher) scale. In this case,
specific manifestations of these states both at colliders
(long-lived squarks and /or sleptons, changes in their decay modes
hierarchy etc.), and in neutralino-nucleon scattering (large
contribution to SI interaction) should be. In this paper, however,
we concentrate only on the scenario with $M_{1/2}\sim M_0$ where
superscalars are very heavy.

Neutralino-nucleon interaction in the scenario behaves as a
threshold process due to the formalism accepted -- the corresponding
term in the Lagrangian is nondiagonal in neutralino fields:
$i(g_2/2\cos\theta)Z_\mu\bar\chi^0_{1}\ga^\mu \chi^0_{2}$. In the
case of pure neutralino states, zero order contributions to $\chi -
N$ interaction correspond to the spin-independent (SI) inelastic
process. This conclusion results from the interaction Lagrangian --
nondiagonal Higgsino current $\chi^0_1\gamma _{\mu}\chi^0_2$
interacts with $Z_{\mu}$ as a vector, rather than an axial vector,
it is a consequence of the real 4-dimensional Majorana formalism
used (see~\cite{28}).

At the tree level the zero order SI cross section for $\chi^0_1-N$ reaction
takes the following threshold form
\begin{eqnarray}
\sigma^{SI}_{\chi^0 N}=\frac{g^4M_N^2}{64\pi \cos^4\theta_W
M_Z^4}\left(\frac{E_N-\Delta M_{\chi^0}}{E_N}\right)^{1/2}. \label{Del}
\end{eqnarray}
In the non-relativistic case $E_N=W_k(m_N/M_{\chi}),$ where $W_k$ is
an average kinetic energy of the neutralino in the Sun neighborhood,
$W_k=M_{\chi}v^2_r/2$. For $M_{\chi^0} \sim 1 \;\mathrm{TeV}$ this
energy $W_k \sim 1\;\mathrm{MeV}.$ So for $\Delta M_{\chi^0}$ as
large as $(1-10)\, \mathrm{GeV}$ the threshold value for the
reaction energy is unattainable. Then the process is forbidden,
and cosmic neutralinos cannot be detected in the direct terrestrial
experiments.

The cross section for the neutralino-nucleon scattering with chargino
production (recharge process) is similar to (\ref{Del}) and for large
$\Delta M_{\chi_c} \approx 0.5\,\Delta M_{\chi^0}$ this channel is
also closed.

Corrections induced by a nonzero mixing and/or loop diagrams cannot
make these nondiagonal reactions visible because part of correction
amplitudes is damped in the limit of pure Higgsino; contributions
from squark exchanges are small due to large squark masses (see
also~\cite{36,37}). Furthermore, due to Majorana nature of
neutralino, nonzero loop contributions are proportional to the small
parameter $q^2/M_Z^2 \ll 1$. As concerns elastic (diagonal)
channels, they contribute to the $\chi-N$ cross section via loops or
due to a nonzero mixing, so their yield is small as well.

Returning to some RG arguments, it seems that when $M_0\gtrsim M_{1/2}$ we can
expect $\Delta M_{\chi}$ somewhat lower than for the case $M_{1/2} \gg M_0$
when the splitting between $\tilde t_1$ and $\tilde t_2$ can be larger.
Certainly, the relative splitting and mixing of these states are really unknown.

So let us consider the second case with $\Delta M_{\chi}< 1 \,\mathrm{GeV}$
and it can be as low as $\sim (100 - 300) \,\mathrm{MeV}$ if mass splittings
are mainly determined by tree contributions. As to collider signature, in this case only photon,
neutrino pair or low energy $e^-e^+,\,\mu^-\mu^+,\,\pi^-\pi^+$ pairs can be created in the final
states, but it is hard to select these events from the background
(see also~\cite{9,10,11}). So in the case of low $\Delta M_{\chi}$, degenerated
Higgsino and chargino are indeed invisible at the LHC.

The Higgsino-nucleon nondiagonal interaction takes place again in
this case, contributing to the SI cross section. Comparing the model
predictions for $\chi-N$ reaction with experimental restrictions on
the SI cross section~\cite{38,39} it is possible to estimate the
$M_{SUSY}$ value. From the inequality
$$\Delta M_{\chi^0} =
(M_Z^2/(M_1M_2)(M_1\cos^2(\theta_W)+M_2\sin^2(\theta_W)<
W_k(M_N/M_{\chi}),$$ it follows that the process $\chi^0_1N \to
\chi^0_2 N'$ is closed when $M_{SUSY} \le 8.3 \cdot 10^{9}
\;\mathrm{GeV}$ and $M_{SUSY} \le 1.2 \cdot 10^{10} \;\mathrm{GeV}$
for $M_{\chi^0_1} = 1.0 \;\mathrm{TeV}$ and $1.4 \;\mathrm{TeV}$,
respectively. These estimations depend on $\tan \beta$ weakly and
are in agreement with the ones given by the RG analysis. So we
conclude that in this scenario the SI inelastic Higgsino-nucleon
scattering cannot be observed experimentally today.

It seems that the inverse inelastic reaction $\chi^0_2N \to\chi^0_1
N'$ is possible, but $\chi^0_2$ states are unstable, so they decayed
a long time ago. In other words, the only inequality $\tau
_{\chi}\leq T_0$ takes place, where $T_0$ is the age of the
Universe. An upper estimation for $\tau_{\chi}$ follows from the
width $\Gamma(\chi_2\rightarrow \chi_1 \nu\bar{\nu})$ (when $\Delta
M_{\chi^0}$ is reasonably small we consider only the most "soft"
channel) and we have
\begin{equation}
 \Gamma_{\chi}=\frac{G^2_F}{60\pi^3}\Delta M^5_{\chi}.
\end{equation}
Then the following restriction emerges
\begin{equation}
 M_{SUSY}\leq M^2_Z\left(\frac{G^2_F T_0}{60\pi^3}\right)^{1/5}.
\end{equation}
With the value $T_0=3.15\cdot 10^{17}\, s$ we get $M_{SUSY}\leq 4.25\cdot 10^9\,\mathrm{GeV}$.
It is again in accordance with the RG results and slightly more stringent
than the restrictions following from the threshold inequality
$\Delta M_{\chi}\geq W_k(M_N/M_{\chi})$.

Electroweak corrections to the splitting $\Delta M_{\chi^{\pm}}$ can be
as small as $\sim 100\, \mathrm{MeV}$ due to loops with $\gamma,
Z$ and $W$~\cite{14}. When $\Delta M_{\chi^{\pm}}\sim 100 \,
\mathrm{MeV}$ (squark loop contributions are too small), recharge
process $\chi^0_1n \to \chi^{\pm}_1p $ accompanied by a track of
$\chi^{\pm}$ seems as possible due to very energetic neutralinos,
but their cross section is strongly damped and the reaction is entirely exotic.
So the recharge process cannot be detected experimentally too.

Depending on mass splitting the chargino lifetime can be estimated in
the following manner: from the channel $\chi^{\pm}\to \chi^0 e \bar
\nu_e$ we get
$$ \tau_{\chi^{\pm}}=(30 \pi^3/G_F^2)\Delta
M_{\chi_c}^{-5},$$ but in the intermediate interval of $\Delta
M_{\chi^{\pm}} =(0.1 - 1.0) \;\mathrm{GeV}$ there are also chargino
decay channels with the final $\mu$ and $\pi$-meson. For these
decays the corresponding formulae are more cumbersome, and we do not
write them here. Gathering all contributions, we get approximately
$$ \tau_{\chi^{\pm}}\sim (10^{-7} - 10^{-12}) \;s.$$

Thus, for both the cases -- large or small $\Delta M_{\chi}$ -- the
subscenarios of this class do not produce practically any visible
signal in various channels of neutralino-nucleon scattering at
modern measuring tools.

If $\Delta M_{\chi}$ values are sufficiently large, there is a
chance to discover at the LHC some decay modes of the lowest
neutralino and chargino states. Moreover, if some specific squark
(slepton) effects occur at low TeV scale too, but gaugino
manifestations are absent, it may correspond to the subscenario
$M_{1/2}\gg M_0\sim \mu$. The existence of specific
neutralino-chargino decays together with the absence of other SUSY
states at the TeV scale can be understood in the framework of the
subscenarios $M_0\gtrsim M_{1/2}\gg \mu$ or $M_{1/2}\gg M_0\gg \mu$.

Absence in an experiment of various decay modes manifestations can
indicate that the subscenarios with $M_0\gtrsim M_{1/2}\gg \mu$ or
$M_{1/2}\gg M_0\gg \mu$ are realized, providing the lowest
neutralino and chargino with a well degenerate spectrum that cannot
be observed in experiment.

Note, for all scenarios with a large splitting between $M_{1/2}$ and
$M_0$ a large contribution to the SI neutralino-nucleon
cross section is possible due to the squark exchange, especially if
the superscalar scale is closed to $\mu$. Naturally, a study of collider
and $\chi-N$ data correlations is necessary, making details of the state
spectrum more precise. Data on neutralino annihilation photon spectrum are
also needed to complete an analysis of capabilities of the scenarios.

\subsection{Diffuse gamma spectrum from the Galactic halo}

As it follows from above, in the Split Higgsino scenario obvious
manifestations of SUSY can be in some latent form: direct
interaction of neutralino with nuclei is too small and it is very
questionable whether effects of degenerate neutralino and chargino
can be detected at the LHC if $\Delta M_{\chi} < 10\, \mathrm{GeV}$.
It seems that a chance to verify this scenario is to study the
photon spectra from neutralino annihilation in the Galactic halo.
The process can produce gamma quants in two ways: direct photon
creation through loop diagrams or formation of a diffuse
(continuous) gamma spectrum due to secondary photons created in
radiative decays of mesons.

In this scenario the dominant mode of continuous gamma spectrum creation
is Higgsino annihilation into $WW$ and $ZZ$ bosons followed by
creation and radiative decays of light mesons, mainly through the
channel $\pi^0 \to 2\gamma$. For the lowest Higgsino mass $M_{\chi}=
1.4\;\mathrm{TeV}$ we calculate the cross section of Higgsino
annihilation into $WW$ and $ZZ$  and get $(\si v)_{WW+ZZ}\approx
0.7\cdot 10^{-26}\;cm^3s^{-1}$. Further, the same spectrum
$$\displaystyle \frac{dN_{\gamma}}{dE_{\gamma}}\approx \frac{0.73}{m_{\chi}}
\frac{e^{-7.76x}}{x^{1.5}},$$ where $x=E_{\gamma}/M_{\chi}$, is used
for both (W and Z) channels approximately (~\cite{40,41}).

Then the total diffuse photon flux from halo can be calculated as
\begin{eqnarray}
\displaystyle \Phi^{\gamma}(E_0,E_m)\approx 9.3 \cdot 10^{-13}\,cm^{-2}\, s^{-1}
\int^{E_m}_{E_0}dE_{\gamma}\frac{dN_{\gamma}}{dE_{\gamma}}\cdot\\ \nonumber
\frac{<\si v>_{WW+ZZ}}{10^{-26}\,cm^3\, s^{-1}}\cdot \left(\frac{1\,
\mathrm{TeV}}{m_{\chi}}\right)\cdot \bar J(\Delta \Omega)\cdot \Delta \Omega,
\label{spectr}
\end{eqnarray}
where $E_0$ is the threshold photon energy for an apparatus, $E_m$
is the maximal registered photon energy, $\bar J(\Delta \Omega)$ is
averaged over angle value of the integral $J(\Psi)$ which contains
information on the DM distribution in halo~\cite{41,42}.

Fixing some value $\bar J(10^{-3})\approx 1.2\cdot 10^3$ that is
typical of the Navarro-Frenk-White profile, where $\Delta\Omega
=10^{-3} \,sr$ is used (see~\cite{42,43}), we evaluate the total
continuous gamma flux that can be measured at space-based telescopes
(EGRET or GLAST) or at ground based Atmospheric Cherenkov Telescopes
(ACT) (HESS, in particular). Then from (\ref{spectr}) for the total
flux we get
\begin{equation}
\begin{array}{l}
\displaystyle  \Phi^{\gamma}_{EGRET}\approx 0.17\cdot 10^{-10}\,cm^{-2}\, s^{-1},
\quad E_0=1\, \mathrm{GeV},\,\,E_m=20\, \mathrm{GeV},
\\[3mm]
\displaystyle  \Phi^{\gamma}_{GLAST}\,\approx 0.19\cdot
10^{-10}\,cm^{-2}\, s^{-1}, \quad E_0=1\, \mathrm{GeV},\,\,E_m=300\,
\mathrm{GeV},
\\[3mm]
\displaystyle  \Phi^{\gamma}_{HESS}\;\;\approx 0.82\cdot
10^{-12}\,cm^{-2}\, s^{-1}, \quad E_0=60\, \mathrm{GeV},\,\,E_m=1\,
\mathrm{TeV}.
\end{array}
\label{dat}
\end{equation}
As it is seen, the calculated values are beyond experimental possibilities of these telescopes
-- at present, only GLAST has some chance to measure the total continuous gamma flux,
because it can detect a gamma flux as small as
$\Phi^{\gamma}_{GLAST}(\mathrm{exp}) \approx 10^{-10}\,cm^{-2}\, s^{-1}$. In
the near future, however, ACT like HESS, for example, will be
able to fix the flux due to high sensitivity level~\cite{41}
$\Phi^{\gamma}_{HESS}(\mathrm{exp}) \approx 10^{-14}\,cm^{-2}\, s^{-1}$. Note
also that similar results were derived in~\cite{8} for direct
photon signals.

A characteristic value of the flux is not characteristic feature of
the scenario, nearly the same values arise in all schemes with
highly degenerate Higgsino as the lowest state. Nevertheless, from
comparison with other model predictions (MSSM, mSUGRA, etc., see,
for example~\cite{41,42,43,44,45,46,47}) we note that channel of
neutralino annihilation into quarks increases the total flux up to
one order of magnitude, so the flux could be well over the GLAST
sensitivity threshold. Then the Split Higgsino scenario prediction
for the diffuse gamma flux can be discriminated from predictions of
models where there is a large contribution of quark and/or Higgs
annihilation modes. Nevertheless, some conclusion on the Split
Higgsino scenario realizability can be made only from the whole data
analysis, using $\chi-N$ scattering, collider experiments, and
photon spectrum data together.

\section{Conclusions}
We have pointed out that from the one-loop RG analysis of the SUSY
$SU(5)$ theory there can be extracted a few sets of energy scales
which are compatible with conventional ideas on the DM structure and
manifestations. Threshold corrections, induced by heavy states near
$M_{GUT}$ -- $M_{24}, \, M_5$, -- are especially important for
establishment of the hierarchies. Due to a specific form of the RG
equations the scalar scale $M_0$ remains arbitrary, and it occurs
the variety of scenarios divided in the two classes: $M_{1/2}\gg
\mu$ or $M_{1/2}\ll \mu$. Note also that the refined RG analysis
with two-loop $\beta$-functions and mass spectrum improved by
radiative corrections can make the energy scale set more precise.
Namely, the $M_0$ scale can be split to establish squark and slepton
scales separately, while the energy scale hierarchies as two global
classes should remain.

In this paper, the hierarchy $M_0\gtrsim M_{1/2} \gg \mu$ (the Split
Higgsino model) was considered in detail. In this case, the
renormalization group approach determines the SUSY breaking scale as
$M_{SUSY} \sim 10^8 - 10^9\,\mathrm{GeV}$.

Due to the degenerate mass spectrum of the lightest states in the
model the coannihilation channels are essential for the effective
annihilation cross section. With the calculated value of
$<\si_{eff}v>$ the relic abundance value is provided by the lowest
Higgsino mass in the interval $1.2\,-\,1.6\,\mathrm{TeV}$.

Experimental observation of this scenario effects crucially depends
on $\Delta M_{\chi^0}$ and $\Delta M_{\chi_c}$. If these differences
are $\sim 10 \, \mathrm{GeV}$, products of $\chi^0_2$ and
$\chi^{\pm}$ decays can be, in principle, detected at the LHC, for
small splittings these decay modes are invisible. At the same time,
if superscalars are close to the lowest Higgsino scale, their
specific manifestations are possible too. In particular, such
squarks at a TeV scale should increase significantly the SI
neutralino-nucleon cross section. Heavy scalars from the hierarchy
$M_0\gg \mu$ do not change the tree level $\chi-N$ cross section
significantly.

Even if there are signals from neutralino and /or chargino decays,
they are hardly detected. Despite these (possible) effects the Split
Higgsino model is characterized by $\chi-N$ scattering with the
small and therefore yet unregistered SI cross section and some
typical value of a continuous annihilation gamma flux. For small
$\Delta M_{\chi}$ values the $M_{SUSY}$ scale is evaluated as
$\lesssim 10^{10} \, \mathrm{GeV}$ in agreement with the RG results
leading to unobserved SI $\chi-N$ scattering.

Conclude, to detect some footsteps of the scenario, it is necessary
to analyze correlation of all collider data, $\chi -N$ cross section
measurements and value of diffuse gamma flux from halo (or direct
photon spectrum). Only a comparison of all measured characteristics
could give some conclusions on the scenario realization. In a sense,
this model presents a class of ``Hidden SUSY'' scenarios which do
not reject SUSY ideas and at the same time, can explain (possible)
absence of obvious SUSY signals at the LHC.

If, however, the neutralino annihilation induced continuous gamma
flux is not detected at all, simultaneously with the absence of
neutralino and chargino decay modes at the LHC, it will mean that
the DM origin cannot be understood in the MSSM framework. Then to
explain the DM origin and properties there should be attracted some
other sources, as gravitino, for example.


\begin{thebibliography}{100}

\bibitem{1}
N.~Arkani-Hamed, S.~Dimopoulos, JHEP 0506, 073 (2005).

\bibitem{2}
N.~Arkani-Hamed, S.~Dimopoulos, G.~F.~Giudice, A.~Romanino, Nucl.
Phys. B \textbf{709}, 3 (2005).

\bibitem{3}
G.~F.~Giudice, A.~Romanino, Nucl. Phys. B \textbf{699}, 65
(2004).

\bibitem{4}
R.~Mahbubani, L.~Senatore, Phys. Rev. D \textbf{73}, 043510 (2006).

\bibitem{5}
A.~Masiero, S.~Profumo, P.~Ullio, Nucl. Phys. B\textbf{712}, 86
(2005) (ArXiv: hep-ph/0412058).

\bibitem{6}
M.~Masip, I.~Mastromatteo, Phys. Rev. D \textbf{73}, 015007 (2006).

\bibitem{7}
M.~Drees, ArXiv: hep-ph/0501106.

\bibitem{8}
G.~Vereshkov, V.~Kuksa, V.~Beylin, R.~Pasechnik,
ArXiv: hep-ph/0410043.

\bibitem{9}
K.~Cheng, C.-W.~Chiang, J.~Song, JHEP, 0471 (2006).

\bibitem{10}
K.~Cheng, J.~Song, Phys. Rev. D \textbf{72}, 055019 (2005).

\bibitem{11}
U.~Chattopadhyay, D.~Choudhury, M.~Drees, P.~Konar, D.P.~Roy, ArXiv:
hep-ph/0508098.

\bibitem{12}
C.S.~Chen, M.~Drees, J.F.~Gunion, Phys. Rev. D \textbf{55}, 330
(1997); Phys. Rev. Lett. \textbf{76} 2002 (1996).

\bibitem{13}
H.~C.~Cheng, B.~Dobrescu, K.~Matchev, ArXiv: hep-ph/9811316;
D.~Hooper, C.-T.~Wang, Phys. Rev. D \textbf{69}, 035001 (2004);
W.~Kilian, T.~Plehn, P.~Richardson, E.~Schmidt, Eur. Phys. J. C
\textbf{39}, 229 (2005); B.~Mukhopadhyaya, S.~Sen Gupta, Phys. Rev. D
\textbf{71}, 035004 (2005); A.~Pierce, Phys. Rev. D \textbf{70},
075006 (2004); S.~Profumo, C.~E.~Yaguna, Phys. Rev. D \textbf{70},
095004 (2004).

\bibitem{14}
D.~Pierce, A.~Papadopoulos, Phys. Rev. D \textbf{50}, 565 (1994);
Nucl. Phys. B \textbf{430}, 278 (1994); A.~B.~Lahanas, K.~Tamvakis,
N.~D.~Tracas, Phys. Lett. B \textbf{324}, 387 (1994); D.~Pierce,
J.~Bagger, K.~Matchev, R.~Zhang, Nucl. Phys. B \textbf{491}, 3
(1997).

\bibitem{15}
W.~\"{O}ller, H.~Eberl, W.~Majerotto, C.~Weber, Eur. Phys. J. C
\textbf{29}, 563 (2004); W.~Majerotto, ArXiv: hep-ph/0209137.

\bibitem{16}
G.~F.~Giudice, A.~Pomarol, Phys. Lett. B \textbf{372}, 253 (1996).

\bibitem{17}
R.~Arnowitt, P.~Nath, Arxiv: hep-ph/9309277.

\bibitem{18}
J.L.~Lopez, D.V.~Nanopoulos, A.~Zichichi, Prog. Part. Nucl. Phys.
\textbf{33}, 303 (1994) (ArXiv: hep-ph/9402299).

\bibitem{19}
J.~Ellis, S.~Kelley, D.~V.~Nanopoulos, Nucl. Phys. B \textbf{737}, 55 (1992);
Phys. Lett. B \textbf{287}, 95 (1992).

\bibitem{20}
J.~Hisano, H.~Murayama, T.~Yanagida, Phys. Rev. Lett. \textbf{69}, 1014 (1992).

\bibitem{21}
P.~Langacker, N.~Polonsky, Phys. Rev. D \textbf{52}, 3081 (1995).

\bibitem{22}
J.~Bagger, K.~Matchev, D.~Pierce, Phys. Lett. B \textbf{348}, 443 (1995).

\bibitem{23}
A.~Dedes, A.~B.~Lahanas, J.~Rizos, K.~Tamvakis, Phys. Rev. D
\textbf{55}, 2955 (1997); J.~A.~Bagger, J.~L.~Feng, N.~Polonsky, R.-J.~Zhang,
ArXiv:hep-ph/9911255; N.~Haba, N.~Okada, ArXiv: hep-ph/0602013;
C.~S.~Aulakh, S.~H.~Carg, ArXiv: hep-th/06012021.

\bibitem{24}
V.~Beylin, G.~Vereshkov, V.~Kuksa, in Proc. of 16 Int. Workshop on
QFT and HEP, Moscow, 2001, p.300.

\bibitem{25}
Review of Particle Properties: S.~Eidelman et al, Phys. Lett. B \textbf{592}, 1 (2004).

\bibitem{26}
S.~Mizuta, M.~Yamaguchi, Phys. Lett. B \textbf{298}, 120 (1993).

\bibitem{27}
M.~Drees, M.~M.~Nojiri, D.~P.~Roy, Y.~Yamada, Phys. Rev. D \textbf{56}, 276 (1997).

\bibitem{28}
OPAL Collaboration: G.~Abbiendi et al, preprint CERN - EP/2002 - 063;
P.~Abreu et al, Eur. Phys. J. C \textbf{1}, 1 (1998);
L3 Collaboration: G.~Grenier in Int. Conf. on SUSY in Physics, SUSY2000, CERN, 2000.

\bibitem{29}
V.~Beylin, V.~Kuksa, R.~Pasechnik, G.~Vereshkov, ArXiv: hep-ph/0702148.

\bibitem{30}
N.~Arkani-Hamed, A.~Delgado, G.~F.~Giudice, Arxiv: hep-ph/0601041.

\bibitem{31}
K.~Griest, M.~Kamionkowski, M.~S.~Turner, Phys. Rev. D
\textbf{41},3565 (1990); K.~A.~Olive, M.~Srednicki, Phys. Lett. B
\textbf{230}, 78 (1989); Nucl. Phys. B \textbf{355}, 208 (1991);
J.~Ellis, J.~C.~Hagelin, D.~V.~Nanopoulos, K.~Olive, M.~Srednicki,
Nucl. Phys. B \textbf{238}, 453 (1989).

\bibitem{32}
 K.~Griest, D.~Seckel, Phys. Rev. D \textbf{43}, 3191 (1991).

\bibitem{33}
G.~Jungman, M.~Kamionkowski, K.~Griest, Phys. Rep. \textbf{267}, 195
(1996).

\bibitem{34}
J.~Edsjo, P.~Gondolo, Phys. Rev. D \textbf{56}, 1879 (1997).

\bibitem{35}
V.~A.~Bednyakov, H.~V.~Klapdor-Kleingrothaus, E.~Zaiti, Phys. Rev. D
\textbf{66}, 015010 (2002).

\bibitem{36}
D.~N.~Spergel, et al., Astrophys. J. Suppl. \textbf{148}, 175
(2003).

\bibitem{37}
J.~Hisano, S.~Matsumoto, M.M.~Nojiri, O.~Saito, Phys. Rev. D
\textbf{71}, 015007 (2005).

\bibitem{38}
M.~Cirelli, N.~Fornengo, A.~Strumia, ArXiv: hep-ph/0512090.

\bibitem{39}
D.~S.~Akerib et al, ArXiv: hep-ph/0509209; Y.~G.~Kim, T.~Nihei,
L.~Roszkowski, R.~Ruiz de Austri, JHEP, \textbf{12}, 034 (2002);
V.~A.~Bednyakov, H.~V.~Klapdor-Kleingrothaus, Phys. Rev. D
\textbf{63}, 095005 (2001);
A.~Djouadi, M.~Drees, Phys. Lett. B \textbf{484}, 183 (2000).

\bibitem{40}
A.~Masiero, S.~Profumo, P.~Ullio, Nucl. Phys. B \textbf{712}, 86
(2005).

\bibitem{41}
L.~Bergstrom, P.~Ullio, J.~H.~Buckley, Astropart. Phys. \textbf{9},
137 (1998).

\bibitem{42}
J.~Hisano, S.~Matsumoto, M.~M.~Nojiri, O.~Saito, Phys. Rev. D
\textbf{71}, 063528 (2005).

\bibitem{43}
Y.~Mambrini, C.~Munoz, ArXiv: hep-ph/0407158.

\bibitem{44}
A.~Cesarini, F.~Fucito, A.~Lionetto, A.~Morselli, P.~Ullio,
ArXiv: astro-ph/0305075.

\bibitem{45}
N.~Fornengo, L.~Pieri, S.~Scopel, Phys. Rev. D \textbf{70}, 103529
(2004) (ArXiv: hep-ph/0407342).

\bibitem{46}
W.~de Boer, M.~Herold, C.~Sandez, V.~Zhukov, Eur. Phys. J. C
\textbf{33} 981 (2004) (ArXiv: hep-ph/0309029).

\bibitem{47}
D.~Hooper, B.~Dingus, ArXiv: astro-ph/0212509.

\bibitem{48}
A.~Provenza, M.~Quiros, P.~Ullio, ArXiv: hep-ph/0609059;
Y.~Mambrini, C.~Munoz, J. Cosm. Astropart. Phys. \textbf{10}, 003
(2004).

\end{thebibliography}
\end{document}